\documentclass[prd,floats,superscriptaddress,showpacs,nofootinbib,twocolumn,preprintnumbers]{revtex4}%

\usepackage{amssymb}
\usepackage{amsmath}

\usepackage{graphicx}
\usepackage{graphics}
\usepackage{dcolumn}
\usepackage{color}
\usepackage{rotate}
\usepackage{fancyhdr} 
\usepackage{hyperref}
\usepackage{indentfirst}

\usepackage{wasysym}

\usepackage{umoline}
\usepackage{ulem}

\def\be{\begin{eqnarray}}
\def\ee{\end{eqnarray}}
\def\no{\nonumber}

\definecolor{darkred}{rgb}{.743,0,0}

\begin{document}
\preprint{CERN-TH-2019-109}
\title{Is there a supernova bound on axions?}

\author{Nitsan Bar}
\email{nitsan.bar@weizmann.ac.il}
\affiliation{Weizmann Institute of Science, Rehovot, Israel 7610001}
\author{Kfir Blum}
\email{kfir.blum@cern.ch}
\affiliation{Weizmann Institute of Science, Rehovot, Israel 7610001}
\affiliation{Theory department, CERN, CH-1211 Geneve 23, Switzerland}
\author{Guido D'Amico}
\email{guido.damico@cern.ch}
%\affiliation{Theory department, CERN, CH-1211 Geneve 23, Switzerland}
\affiliation{Stanford Institute for Theoretical Physics, Stanford University, Stanford, CA 94306, USA}

%\date{\today}

\begin{abstract}
We present a critical assessment of the SN1987A supernova cooling bound on axions and other light particles. Core collapse simulations used in the literature to substantiate the bound omitted from the calculation the envelope exterior to the proto-neutron star (PNS). As a result, the only source of neutrinos in these simulations was, by construction, a cooling PNS. We show that if the canonical delayed neutrino mechanism failed to explode SN1987A, and if the pre-collapse star was rotating, then an accretion disk would form that could explain the late-time ($t\gtrsim5$~sec) neutrino events. Such accretion disk would be a natural feature if SN1987A was a collapse-induced thermonuclear explosion. Axions do not cool the disk and do not affect its neutrino output, provided the disk is optically-thin to neutrinos, as it naturally is. These considerations cast doubt on the supernova cooling bound.
\end{abstract}
%\pacs{95.35.+d,98.35.Gi}

\maketitle

%\tableofcontents

%%%%%%%%%%%%%%%%%%%
\section{Introduction}\label{sec:intro}

The neutrino burst of the core collapse supernova (CCSN) SN1987A~\cite{Hirata:1987hu,Bionta:1987qt,Alekseev:1987ej} played an important role in constraining models of new light particles beyond the Standard Model. A good example is the Peccei-Quinn axion~\cite{Peccei:1977hh,Wilczek:1977pj,Weinberg:1977ma}, a target of extensive experimental searches~\cite{Raffelt:1990yz,Raffelt2008,Tanabashi:2018oca,Armengaud:2019uso}. 
In this paper we present a critical reassessment of the SN1987A bound on new particles. We highlight the case of axions, but we expect that our discussion also applies to other models such as dark photons~\cite{Chang2017}, sterile neutrinos~\cite{Arguelles:2016uwb}, and other examples~\cite{Fuller:1988ega,Choi1990,Hanhart2001,Chang2018a}.

\begin{figure}[htbp]
\begin{center}
\includegraphics[width=0.475\textwidth]{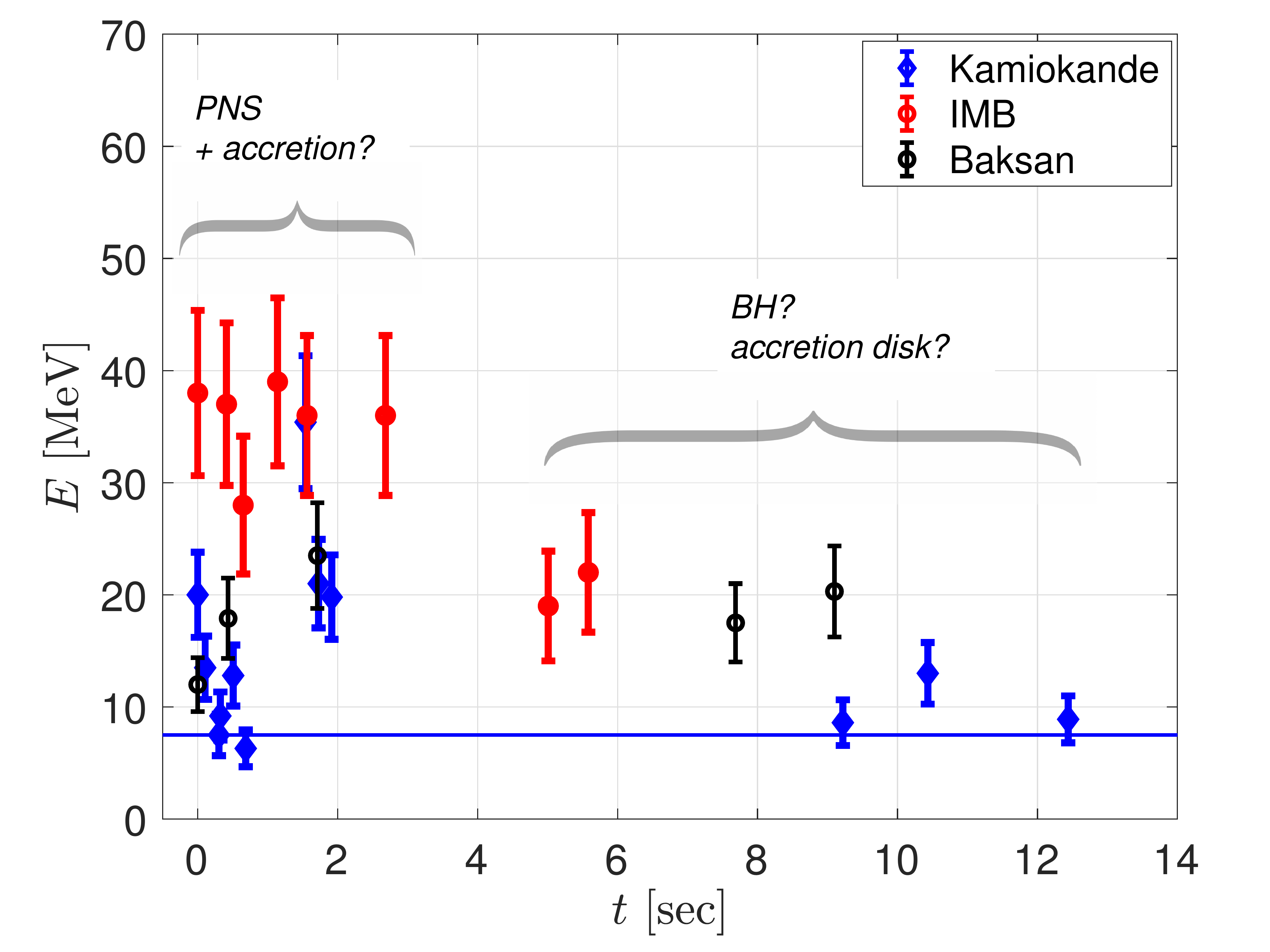}
\includegraphics[width=0.49\textwidth]{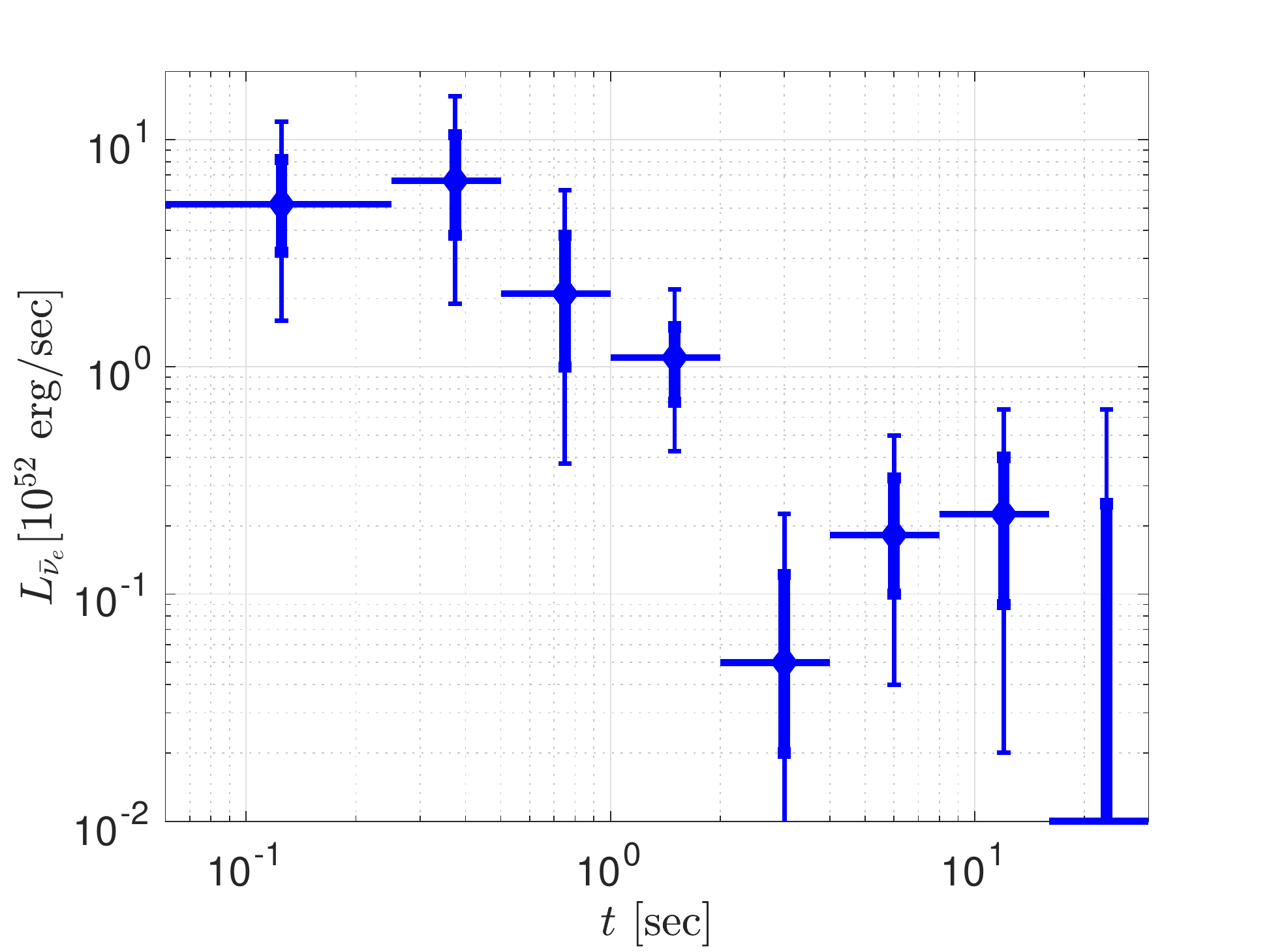}
\caption{{\bf Top:} SN1987A neutrino event energies and arrival times as registered by the detectors Kamiokande~\cite{Hirata:1987hu}, IMB~\cite{Bionta:1987qt}, and Baksan~\cite{Alekseev:1987ej}. Horizontal line indicates the energy threshold adopted in the Kamiokande analysis~\cite{Hirata:1987hu}. 
{\bf Bottom:} Inferred $\bar\nu_e$ luminosity at the source. Thick (thin) error bars denote the $1\sigma$ ($2\sigma$) allowed range. Based on analysis in~\cite{Blum:2016afe}. 
}
\label{fig:data}
\end{center}
\end{figure}

The SN1987A neutrino data is reproduced in Fig.~\ref{fig:data}.
The axion bound revolves around the duration of the neutrino signal~\cite{Raffelt:1990yz}. 
The argument is that emission of new particles from the proto-neutron star (PNS) core would compete with the Standard Model neutrino production. As a result, the PNS cools too fast to account for the neutrino events observed up to $\sim10$~sec after the onset of the burst. 

This argument 
for the axion bound 
hinges on the assumption that the late-time ($t\gtrsim5$~sec) neutrino emission was produced by a cooling PNS. 
But how certain are we that this was the case? 
We are concerned about the following issues:
\begin{enumerate}
\item A cooling PNS is not the only source of neutrino emission in CCSNe. If matter from the envelope of the star continues to accrete onto the core, then accretion contributes to the neutrino luminosity. As we show in Sec.~\ref{sec:prev}, none of the simulations quoted with respect to the axion bound~\cite{Raffelt:1990yz} included the stellar envelope outside of the PNS, so these simulations missed the late-time emission of neutrinos from the accretion of that part of the envelope. As we show in Secs.~\ref{sec:disk} and~\ref{sec:axdonot}, axions do not affect the accretion-induced neutrino luminosity. 
\item The reason that past works focused on a bare PNS, is that a PNS is thought to be the likely object remaining behind in the core of a CCSN that explodes via the delayed neutrino mechanism (D$\nu$M)~\cite{Bethe:1984ux,Kotake:2005zn,Burrows:2012ew,Janka2012}, and the D$\nu$M was assumed to be the cause of the explosion in SN1987A. However, D$\nu$M simulations have not yet been able to reproduce an explosion with progenitor or energetics resembling SN1987A. Success could be around the corner~\cite{Vartanyan:2018xcd}, but we think it important to keep an open mind to the possibility that the D$\nu$M fails. 
\item Lacking a self-consistent D$\nu$M simulation, derivations of the axion bound used simulations in which the explosion was triggered by hand, if it was triggered at all: most analyses simply considered a bare PNS without worrying about the actual supernova. 
Note that the D$\nu$M is not the only proposed model for CCSNe~\cite{Janka:2012wk,LeBlanc:1970kg,1979SvA....23..705A,Ardeljan:2004fq,Burrows:2007yx,Takiwaki:2010cf,Burrows:2005dv,Gilkis:2015adr}. At least some CCSNe could be collapse-induced thermonuclear explosions (CITEs), an idea brought up by~\cite{Burbidge1957,Fowler:1964zz,Hoyle:1960zz,Kushnir:2014oca} and demonstrated in 2D simulations by~\cite{Kushnir:2015mca,Kushnir:2015vka}, specifically focusing on SN1987A in~\cite{Blum:2016afe}. 
CITE needs a rotating progenitor, which was almost certainly the case for SN1987A~\cite{Chita:2008gc}. Rotation leads to the formation of an accretion disk that can sustain the required neutrino luminosity at $t\gtrsim5$~sec~\cite{MacFadyen:1998vz,Popham:1998ab,Liu:2015prx,Blum:2016afe}. 
\item The outcome of prolonged accretion is a stellar-mass black hole (BH), rather than a neutron star (NS) remnant. Interestingly, a NS has not been observed in the remnant of SN1987A~\cite{Graves:2005xy,2016ARA&A..54...19M,Esposito:2018nib,Alp:2018oek}. 
This is not necessarily a problem for models predicting a NS as it could still be hiding among the debris of the supernova~\cite{Esposito:2018nib,Alp:2018oek}. Nevertheless, it would be reassuring to see observational evidence for the NS's existence.
\end{enumerate}
We conclude in Sec.~\ref{sec:sum} that the current understanding of SN1987A does not allow to place robust constraints on axions, or other new particles that free-stream out of the supernova core. Experimental searches (e.g. the upcoming IAXO experiment~\cite{Armengaud:2019uso}) would do good to keep an open eye for axions in the parameter space nominally excluded by the supernova bound.

The next Galactic supernova~\cite{Adams:2013ana} could be more conducive to new physics searches via detailed information on the neutrino flavour composition and time structure of the burst~\cite{Bar2019,Mirizzi:2015eza,Fischer:2016cyd,Blum:2016afe}. 
In addition, some credence may be lent to the bound if a NS would be observed in the remnant, or if D$\nu$M simulations would demonstrate explosions with progenitors and energetics comparable to those of SN1987A. However, even then one might worry about residual accretion that could accompany the explosion also in the D$\nu$M~\cite{Mueller:2014rna,Seadrow2018} once 3D effects are taken into account. 

\section{Previous work}\label{sec:prev}

Raffelt~\cite{Raffelt:1990yz,Raffelt2008} reviewed the SN1987A bound on axions and suggested the following criterion to define the axion bound, based on the time duration of the burst:
\be\label{eq:raffelt}\epsilon_a&<&10^{19}~{\rm erg/g/s}.\ee
Here $\epsilon_a$ is the axion emissivity which, in Eq.~(\ref{eq:raffelt}), is to be evaluated at reference temperature $T=30$~MeV and density $\rho=3\times10^{14}$~g/cm$^3$. 
Eq.~(\ref{eq:raffelt}) was introduced as a simple, effective means to formulate the axion bound. To substantiate it, Raffelt~\cite{Raffelt:1990yz} refers to the numerical simulations of Mayle et al~\cite{Mayle:1987as,Mayle:1989yx} and of Burrows, Turner, and Brinkmann~\cite{Burrows1989}. We therefore discuss these numerical works in some detail.

Ref.~\cite{Mayle:1989yx} (updating~\cite{Mayle:1987as}) studied core collapse in 1D simulations including axions. The axion bound was defined to correspond to the value of $f_a$ with which the neutrino burst duration extends over $>7$~sec. This timing requirement implied total energy emission in axions of $E_a=\int dtL_a<3\times10^{53}$~erg. 
Two points are important to note about the simulations of~\cite{Mayle:1987as,Mayle:1989yx}: 
\begin{itemize}
\item There was no supernova explosion in the simulations.
\item The simulations included only the central M=1.64~M$_\odot$ iron core of the star, discarding the stellar envelope outside of it. The time it takes from the onset of core collapse until the outermost mass coordinate of~\cite{Mayle:1987as,Mayle:1989yx} falls onto the PNS is $t\approx
0.4\left(1.64~{\rm M_\odot}/M\right)^{\frac{1}{2}}\left(r/2\times10^8~{\rm cm}\right)^{\frac{3}{2}}$~sec, where $r\approx2\times10^8$ is a typical radial coordinate for this value of M. Therefore, on times $t\gtrsim1$~sec or so, the simulations left out of the calculation the accretion of the envelope outside of the iron core.
\end{itemize}

Inspecting the neutrino luminosity in the calculations of~\cite{Mayle:1989yx} (see Fig.~3 there), one notices that the neutrino luminosity during $t<2.5$~sec is insensitive to axion emission, regardless of the value of $f_a$. At $t>2.5$~sec axion cooling starts to affect the neutrino signal, but by $t=5$~sec the neutrino luminosity is still only reduced by a factor of $\sim2$ compared to the no-axion simulation. Such minor suppression would be perfectly compatible with the SN1987A data (see Fig.~\ref{fig:data}). 

The fact that axions do not affect the early phase of the neutrino burst may seem surprising on first glance. The reason for that, is that while axion emission can quickly drain the PNS core of internal energy, it takes a few seconds (of order the neutrino diffusion time) for this information to propagate to the neutrinosphere where the neutrino signal is determined. We have verified the general behaviour found by~\cite{Mayle:1989yx} using numerical simulations, described in App.~\ref{app:sim}.

At $t>5$~sec, the neutrino luminosity in the simulations of~\cite{Mayle:1989yx} with $f_a$ in the excluded range goes significantly below the no-axion case, falling to $L_{\bar\nu_e}<10^{51}$~erg/s at $t\approx7$~sec. If the PNS was the only source of neutrinos, then this behaviour would indeed be inconsistent with the neutrino events around $t\sim10$~sec.

The second suite of simulations referred to by~\cite{Raffelt:1990yz} is that of Burrows, Turner, and Brinkmann~\cite{Burrows1989}, based on the numerical framework of~\cite{Burrows:1988ba,Burrows:1986me}. Again, the simulations (with~\cite{Burrows1989} and without~\cite{Burrows:1988ba,Burrows:1986me} axion emission) did not involve a supernova explosion. The explicit initial conditions contained only the iron core with a mass of $M=1.3$~M$_\odot$. Ref.~\cite{Burrows1989} added a treatment of accretion, but that was not calculated from an actual stellar profile. Instead, an effective accretion rate was specified by $\dot M= \dot M_0e^{-t/\tau}$. Three models were studied: model A, with $\dot M_0=1$~M$_\odot$/s, and models B and C, with $\dot M_0=0.4$~M$_\odot$/s. All three models used $\tau=0.5$~sec. 
With these parameters the accretion rate in all three models dropped below $10^{-3}$~M$_\odot$/s within less than 2.8 seconds, effectively eliminating the accretion component of the neutrino luminosity that, as we show in Sec.~\ref{sec:axdonot}, requires $\dot M\gtrsim0.05$~M$_\odot$/s to accommodate the SN1987A data. With this treatment, effectively limiting the simulations to contain a bare cooling PNS at $t\gtrsim2$~sec, Ref.~\cite{Burrows1989} found an axion bound that was approximately consistent with the results of~\cite{Mayle:1989yx} (after proper matching of the axion coupling definitions~\cite{Raffelt:1990yz}).

Proceeding from Raffelt's work in the 90's~\cite{Raffelt:1990yz} to more recent analyses, the strategy remained the same: the emission of new particles was calculated in simulations of PNS cooling, without a supernova explosion. Ref.~\cite{Keil:1994sm} simulated a bare PNS read-off from a non-exploding core collapse simulation at $t=0.5$~sec. Ref.~\cite{Fischer:2016cyd} (see also~\cite{Lee:2018lcj}) used simulations in which an explosion was triggered by artificially enhancing the heat deposition due to neutrinos behind the stalled shock. The artificial heating rates were tuned such that by $t=0.5$~sec, the shock progressed out to $\sim1000$~km, thereby eliminating the accretion luminosity component. Refs.~\cite{Chang2017,Chang2018a} used simulations from~\cite{Nakazato:2012qf} in which an explosion was set-off artificially at $t=0.1$~sec, after which the envelope above the PNS was removed by hand.

In summary, the Raffelt criterion~\cite{Raffelt:1990yz} Eq.~(\ref{eq:raffelt}) is based on the assumption that non-exploding D$\nu$M simulations must somehow be missing a key aspect of the physics, such that in reality the D$\nu$M must trigger an explosion on time $t\lesssim2$~sec after core collapse. The explosion is assumed to strip-off the envelope of the star, leaving the cooling PNS as the only source of neutrinos. Following this logic, all of the analyses of the supernova axion bound effectively involved simulations of bare PNS cooling, leaving the rest of the star out of the calculation.  
This scenario {\it could} be correct: perhaps D$\nu$M simulations would eventually achieve self-consistent explosions a-la SN1987A (see, e.g.~\cite{Vartanyan:2018xcd}). In that case, the supernova axion bound could perhaps be substantiated. 

However, if the D$\nu$M failed in SN1987A, then the stellar envelope would have continued to accrete onto the compact central object. In that case one is left to wonder whether the accretion-induced neutrino luminosity could invalidate the axion bound. In the next two sections we attend to this question. In Sec.~\ref{sec:disk} we review the possibility, brought up and studied using numerical simulations in~\cite{Blum:2016afe}, that the late-time events of SN1987A came from an accretion disk. In Sec.~\ref{sec:axdonot} we show that axion emission does not affect the neutrino luminosity of such a disk.

\section{Late-time events from an accretion disk}
\label{sec:disk}
If the D$\nu$M fails to explode the star, then the continued accretion of the envelope could lead to BH formation within $\sim1-3$~sec~\cite{OConnor:2010moj}. BH formation under quasi-spherical accretion would temporarily quench the neutrino luminosity, leading to a quiescent phase lasting a few seconds. As shown in Ref.~\cite{Blum:2016afe}, a gap in the SN1987A neutrino burst, starting around $t\sim2-3$~sec, is consistent with the data. This gap was also noted in~\cite{Spergel:1987ch,Suzuki:1988qi,Lattimer:1989zz}\footnote{In fact Loredo \& Lamb~\cite{Loredo2002}, in their classic phenomenological--statistical analysis of SN1987A neutrinos, noted that their best-fit cooling PNS+accretion model entails BH formation; see their Secs.~VI.A and~VIII.D. Loredo \& Lamb~\cite{Loredo2002} discarded the BH solution because, lacking an alternative hypothesis to the D$\nu$M, they could offer no explanation for late-time events.}.

Next, if the pre-collapse star was rotating, an accretion disk forms. Such accretion disks around stellar-mass BHs were studied in numerical simulations by different groups (see, e.g.~\cite{MacFadyen:1998vz,Popham:1998ab,Liu:2015prx,Blum:2016afe}). 
Because the formation of the disk is associated with the failure of the D$\nu$M to produce an explosion, Ref.~\cite{MacFadyen:1998vz} considered this scenario a ``failed supernova". However, if the CITE model operates~\cite{Kushnir:2014oca,Kushnir:2015mca,Kushnir:2015vka}, then at least some of these ``failed supernovae" may not fail after all. What if a disk formed in SN1987A~\cite{Blum:2016afe}?

The details of disk formation depend on the specific angular momentum, $j$, in the star. Having started to collapse, a mass element located on the rotation plane at mass coordinate $M$ and with specific angular momentum $j$ would hit a centrifugal barrier at a distance
\be\label{eq:Rdisk} R_{\rm disk}&\approx&\frac{j^2}{2GM}\;\approx\;50\left(\frac{j}{5\times10^{16}\rm cm^2/s}\right)^2\left(\frac{2~\rm M_\odot}{M}\right)~{\rm km}\no\\&&
\ee
above the compact remnant. The time at which the mass element reaches $R_{\rm disk}$ can be estimated by~\cite{MacFadyen:1998vz,Kushnir:2014oca,Blum:2016afe}
\be\label{eq:tdisk} t_{\rm disk}&\approx&\pi\sqrt{\frac{r_0^3}{2GM}}
\approx4\left(\frac{r_0}{10^9~{\rm cm}}\right)^{\frac{3}{2}}\left(\frac{2~{\rm M_\odot}}{M}\right)^{\frac{1}{2}}~{\rm sec}\no\\&&\ee
where $r_0$ is the pre-collapse position. Note that $t_{\rm disk}$ equals twice the Keplerian free-fall time for the mass element~\cite{Blum:2016afe} and is increasing with $r_0$.

\begin{figure}[htbp]
\begin{center}
\includegraphics[width=0.5\textwidth]{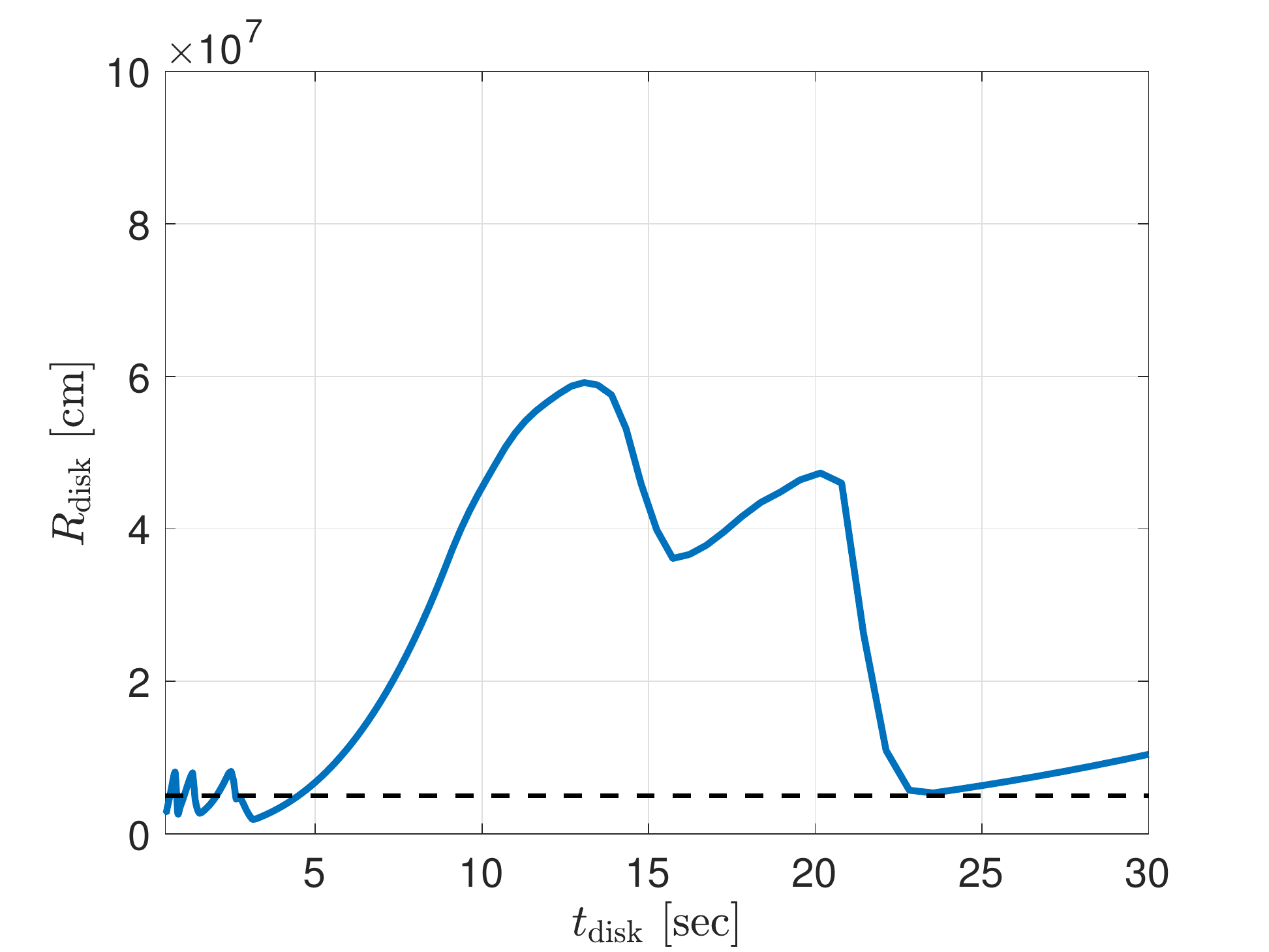}
\caption{Keplerian disk formation radius vs. disk formation time, for the 25~M$_\odot$ rotating stellar model of Hirschi, Meynet\,\&\,Maeder~\cite{Hirschi:2004ks}. The horizontal dashed line denotes a reference radius of 50~km.
}
\label{fig:rdisk}
\end{center}
\end{figure}

An accretion disk forms when the value of $R_{\rm disk}$ emerges above the compact remnant and continues to increase with time. 
While stellar internal rotation profiles are not yet well understood, especially just before core collapse~\cite{Maeder:2000wv,Meynet:2013ih}, the reference values of $R_{\rm disk}$ and $t_{\rm disk}$ in Eqs.~(\ref{eq:Rdisk}-\ref{eq:tdisk}) are indicative of disk formation inferred from some stellar evolution models~\cite{Heger:1999ax,Heger:2003nh,Hirschi:2004ks}. As an example, we use the rotating star model of Hirschi, Meynet\,\&\,Maeder~\cite{Hirschi:2004ks} to extract $R_{\rm disk}$ as a function of $t_{\rm disk}$ for their 25~M$_\odot$ progenitor, and show the result in Fig.~\ref{fig:rdisk}. This model predicts an accretion disk forming $\sim5$~sec after core-collapse with a base radius of $\sim50$~km. 

For a progenitor profile relevant for SN1987A, the mass accretion rate of matter falling through the disk is in the ballpark of $\dot M\sim0.05$~M$_\odot$/s and can be sustained for many seconds~\cite{MacFadyen:1998vz,Popham:1998ab,Liu:2015prx,Blum:2016afe}. 
The accretion rate can be used to estimate the accretion luminosity,%\footnote{An equal luminosity is also emitted in $\nu_e$.},
\be\label{eq:Lacc}
L_{\bar{\nu}_e} &\approx& \frac{GM_{\rm rem}\dot{M}}{2R_{\rm disk}} \\& \approx& 2.6\times10^{51} \bigg(\dfrac{M_{\rm rem}}{2M_{\odot}}\bigg)\bigg(\dfrac{\dot{M}}{0.05M_{\odot}/{\rm s}}\bigg)\left(\dfrac{50\text{km}}{R_{\rm disk}}\right) {\rm erg\,s^{-1}}.\no
\ee
This accretion luminosity is dominated by nucleon conversion ($\beta$ and inverse-$\beta$ decay) with $L_{\nu_e}\approx L_{\bar\nu_e}\gg L_{\nu_x}$ at the source~\cite{Bar2019}.
For an optically-thin disk, the neutrino spectrum approximately follows a pinched Fermi-Dirac spectrum with mean neutrino energy related to the emitting plasma temperature via $\left\langle E_{\bar\nu_e}\right\rangle\approx5.07\,T$. Numerical simulations with $R_{\rm disk}\sim50$~km,  $M_{\rm rem}\sim(2-3)$~M$_\odot$, and $\dot M\sim0.05$~M$_\odot$/s find $T\sim2.5$~MeV~\cite{MacFadyen:1998vz,Blum:2016afe}\footnote{See the snapshot at $t=5.5$~sec in Fig. 5 in~\cite{Blum:2016afe}, and at $t=7.598$~sec in Fig.~6 in~\cite{MacFadyen:1998vz}.}, for which $\left\langle E_{\bar\nu_e}\right\rangle\approx12.7$~MeV. These results for $L_{\bar\nu_e}$ and $\left\langle E_{\bar\nu_e}\right\rangle$ are consistent with the SN1987A data in Fig.~\ref{fig:data}, for the late-time events at $t\gtrsim5$~sec. 

The disk neutrino luminosity can be maintained for a time scale of 10-20~sec. It decays eventually, following the drop in mass accretion rate which, in turn, is sensitive to the initial stellar profile. In the simulations of~\cite{MacFadyen:1998vz}, for example, this drop started around $t\sim15$~sec. Apart from the decreasing mass accretion rate, if CITE operates then thermonuclear detonation cuts-off the accretion luminosity at a time  of the order of the free-fall time of the oxygen layer outer boundary. In the SN1987A CITE simulation of~\cite{Blum:2016afe}, the explosion occurred at $t\approx25$~sec. 
The Kamiokande neutrino data~\cite{Hirata:1987hu} does contain four events between 17 and 24~sec. Considering the total background rate $B\approx0.187$~Hz~\cite{Loredo2002}, these events are consistent with the background; nevertheless, with CITE as an alternative, accretion disk luminosity could conceivably associate some of these events with genuine signal.

\section{Axion emission does not affect accretion disk neutrino luminosity}\label{sec:axdonot}
Axion emission with values of $f_a$ within a few orders of magnitude from the standard axion bound ($f_a\sim10^8$~GeV~\cite{Raffelt:1990yz}) does not affect the neutrino emission of the disk. To see this, we model the axion emissivity by~\cite{Raffelt2008}
\be
\label{eq:epsa}\epsilon_a &\approx& 4.7\times 10^{20}\left(\dfrac{\rho}{10^{14}\text{~g~cm$^{-3}$}}\right)\left(\dfrac{T}{30~\text{MeV}}\right)^{3.5}\,\times\no\\
&&\left(\dfrac{4\times 10^{8}\text{~GeV}}{f_a}\right)^2 \text{erg~g$^{-1}$~s$^{-1}$}.
\ee
The details of the axion couplings are not very important for the discussion.  For concreteness, in Eq.~(\ref{eq:epsa}) we assumed that the dominant axion emission mechanism is nucleon bremsstrahlung $NN\to NNa$ and used $C_N=1$ in the dilute approximation~\cite{Raffelt2008}. 
For comparison, the $\bar\nu_e$ emissivity can be estimated including only nucleon conversion~\cite{Qian:1996xt},
\be\label{eq:epsnu}
\epsilon_{\bar\nu_e} &\approx& 2.7\times 10^{20}\left(\frac{X_n}{0.5}\right)\left(\dfrac{T}{2.5\text{~MeV}}\right)^6\text{erg~g$^{-1}$~s$^{-1}$},
\ee
where $X_n$ is the neutron fraction and we consider the disk to consist of a dissociated plasma of $n,p,e^\pm$. 
Using again characteristic values for the density and temperature consistent with simulations~\cite{MacFadyen:1998vz,Blum:2016afe}  
we see that the axion emissivity of the disk is negligible compared to the neutrino emissivity, for values of $f_a$ within 4 orders of magnitude of the canonical axion bound:
\be\label{eq:epsa2epsnu}
\frac{\epsilon_{\bar\nu_e}}{\epsilon_a} &\approx& 3.4\times 10^8 \left(\frac{X_n}{0.5}\right)\left(\dfrac{\rho}{10^{9}\text{~g~cm$^{-3}$}}\right)^{-1}\left(\dfrac{T}{2.5~\text{MeV}}\right)^{2.5}\,\no\\
&&\times\left(\dfrac{f_a}{4\times 10^{8}\text{~GeV}}\right)^2.
\ee

What makes the disk insensitive to axions is not just the smallness of $\epsilon_a$ compared to $\epsilon_{\bar\nu_e}$, shown by Eq.~(\ref{eq:epsa2epsnu}). Even in the high density core of a PNS, with $\rho\sim10^{14}$~g/cm$^3$ and $T\sim30$~MeV, the axion emissivity is small compared to the neutrino emissivity. Rather, the key point is that the accretion disk emission region is characterised by relatively low density, $\rho\sim10^9$~g/cm$^3$, and consequently it is optically-thin to neutrinos: the mean free path of $\bar\nu_e$ is $l\sim3\times10^3\left(10^9~{\rm g\,cm^{-3}}/\rho\right)\left(10~{\rm MeV}/E_\nu\right)^2$~km, to be compared to a characteristic disk scale of $R_{\rm disk}\lesssim100$~km. Therefore, the  power generated in neutrinos via Eq.~(\ref{eq:epsnu}) flows directly out of the star to form the asymptotic luminosity of Eq.~(\ref{eq:Lacc}), being the dominant cooling mechanism of the plasma in the disk. In contrast, a PNS at $\rho\sim10^{14}$~g/cm$^3$ is deeply optically-thick to neutrinos, cannot cool by neutrino volume emission, and can thus be affected by the volume emission of free-streaming axions even for $\epsilon_a<\epsilon_\nu$.

\section{Discussion and Conclusions}\label{sec:sum}
The explosion mechanism of core collapse supernovae (CCSNe) in general, and SN1987A in particular, is still unknown. Nevertheless, the SN1987A neutrino burst had traditionally been used to place constraints on new light particles, such as axions, that free-stream out of the CCSN core. 

Simulations used in the literature to substantiate the axion bound excised, by hand, the envelope of the star above the proto-neutron star (PNS), such that the only source of neutrinos in these simulations was a bare cooling PNS. 
But a cooling PNS is not the only source of neutrinos in a CCSN. If the delayed neutrino mechanism (D$\nu$M) fails, and if the pre-collapse star was rotating, then an accretion disk forms on a time scale of seconds at a typical radius of a few 10's of km above a stellar-mass compact object (neutron star or black hole). 
Such accretion disk would be a natural feature of collapse-induced thermonuclear explosion (CITE). The accretion disk can explain the late-time ($t\gtrsim5$~sec) neutrino events of SN1987A~\cite{Blum:2016afe}. Axions do not cool the disk and do not affect its neutrino output. 

We would like to emphasise that much work remains before CITE can be verified (or excluded) as the explosion mechanism of all or even some CCSNe. This said, even if one takes for granted the D$\nu$M as the explosion mechanism of SN1987A, simulations of the D$\nu$M in 2D show accretion-induced neutrino luminosity that continues even while the star is exploding~\cite{Mueller:2014rna,Seadrow2018}, in contrast to results in 1D. This suggests that a proper evaluation of the axion bound may require 3D simulations extending to $t\sim10$~sec. Without such simulations it may be difficult to ascertain that the $t\sim10$~sec neutrino events did not come from residual accretion. 

We believe that these considerations cast doubt on the  SN1987A cooling bound on axions. Experimental searches would do good to keep an open eye for axions in the parameter space nominally excluded by the canonical supernova bound. Interestingly, if an axion really does exist with parameters in the ``excluded" range, then there should be a diffuse supernova axion background~\cite{Raffelt:2011ft}.

Our discussion of the bound pertains to the neutrino burst duration argument of~\cite{Raffelt:1990yz}. 
An independent argument that bypasses our criticism for some particle physics models was proposed in~\cite{DeRocco:2019njg}, which noted that dark photons free-streaming from the PNS could convert into Standard Model photons or $e^\pm$ pairs outside of the star, leading to tension with gamma-ray limits. 
Another argument~\cite{Sung:2019xie} notes that new particles must not transfer too much of the internal energy of the core ($\gtrsim10^{53}$~erg) into the kinetic energy of the ejecta ($E_{\rm kin}\sim10^{51}$~erg~\cite{Utrobin:2011da}). This consideration may indeed be more robust to the uncertainties of the explosion mechanism compared with the neutrino burst duration argument. 
One point to note is that the time available for the PNS to inject the new particles could be limited by black hole formation at $t_{\rm BH}\lesssim3$~sec or so~\cite{Blum:2016afe}, compared to the injection time of order 10 seconds assumed in~\cite{DeRocco:2019njg,Sung:2019xie}.

Finally, while we focused on axions for concreteness, we expect that the situation is similar with regards to other feebly-interacting new particles such as Majorons~\cite{Fuller:1988ega,Choi1990}, dark photons~\cite{Chang2017}, sterile neutrinos~\cite{Arguelles:2016uwb}, KK gravitons~\cite{Hanhart2001} and other examples~\cite{Chang2018a}.\\

{\it Note added.} As this paper was being prepared for publication, Ref.~\cite{Page:2020gsx} appeared claiming detection of a NS in the remnant of SN1987A. The observational data discussed in~\cite{Page:2020gsx} is an ALMA detection of a hot dust blob in the remnant~\cite{Cigan:2019shp}. However, as noted in~\cite{Page:2020gsx}, the hot blob could also be hiding an accreting BH, rather than a NS. Moreover, there is no evidence for a pulsar. Thus, the reason that~\cite{Page:2020gsx} claimed a NS is actually not the detection of the hot blob. Instead, the reason is due to the fact that~\cite{Page:2020gsx} assumed the D$\nu$M as the explosion mechanism of SN1987A, as follows.  According to~\cite{Page:2020gsx}, ``explosion models" of SN1987A, quoted from Refs.~\cite{Utrobin:2018mjr} and~\cite{Ertl:2019zks}, indicate a gravitational mass $M$ for the compact remnant that is smaller than the critical mass needed to form a BH (about 2.3~M$_\odot$). This, again according to~\cite{Page:2020gsx}, ``strongly suggests that a BH remnant in SN1987A is unlikely". 
However, the simulations of Ref.~\cite{Utrobin:2018mjr,Ertl:2019zks} were artificial D$\nu$M explosions: the explosion energy and the compact remnant mass in these simulations were free parameters, selected by hand (see Sec.2.2 in~\cite{Utrobin:2018mjr} and Sec.3.1 in~\cite{Ertl:2019zks}) by artificially adjusting the neutrino luminosity of the compact object until explosion is reached. In other words, Ref.~\cite{Page:2020gsx} assumed a NS, based on D$\nu$M artificial explosion simulations that assumed a NS. Needless to say, this logic does not disfavour CITE in any way.

\section*{Acknowledgements}
We thank Deog-Ki Hong, Doron Kushnir, Hyun Min Lee, Alessandro Mirizzi and Amir Sharon for discussions and John Beacom, Thomas Janka, Yosef Nir, Georg Raffelt and Yotam Soreq for helpful comments on the manuscript. 
The work of NB and KB was supported by grant 1937/12 from 
the I-CORE program of the Planning and Budgeting Committee and the Israel
Science Foundation and by grant 1507/16
from the Israel Science Foundation. KB is incumbent
of the Dewey David Stone and Harry Levine career
development chair. GDA is supported by the Simons Foundation Origins of the Universe program (Modern Inflationary Cosmology collaboration).

\clearpage

\begin{appendix}

\section{Axions do not affect the early neutrino burst: numerical simulations}\label{app:sim}
In this appendix we use numerical simulations to reproduce the finding of Refs.~\cite{Mayle:1987as,Mayle:1989yx,Burrows1989}, that even when axion emission quickly drains the PNS core of its internal heat reservoir, the surface emission of neutrinos remains unaffected for a while. Thus the first few seconds of a CCSN neutrino burst are insensitive to axion emission.

We employ the one-dimensional GR1D code~\cite{2010CQGra..27k4103O,OConnor:2014sgn} for a 20~M$_\odot$ zero-age main-sequence progenitor star, with the KDE0v1 equation of state~\cite{Agrawal:2005ix}. Our progenitor profile is taken from the non-rotating, solar metallicity sample of~\cite{Woosley:2002zz}. Note that while the CITE scenario we have in mind for SN1987A requires some stellar rotation, Ref.~\cite{Bar2019}  demonstrated that this rotation is likely irrelevant during the first few seconds after core collapse (see App.~D there), so non-rotating profiles are adequate for our current purpose. We add the axion cooling term Eq.~(\ref{eq:epsa}) to the simulations, ignoring axion re-absorption. This is an approximation that only applies in the free-streaming limit. Our simulations cover only the initial 0.5~sec after core collapse, but that is sufficient to demonstrate the axion-driven volume cooling of the PNS along with the insensitiviety of the neutrino surface emission to that cooling.

\begin{figure}[htbp]
\begin{center}
\includegraphics[width=0.475\textwidth]{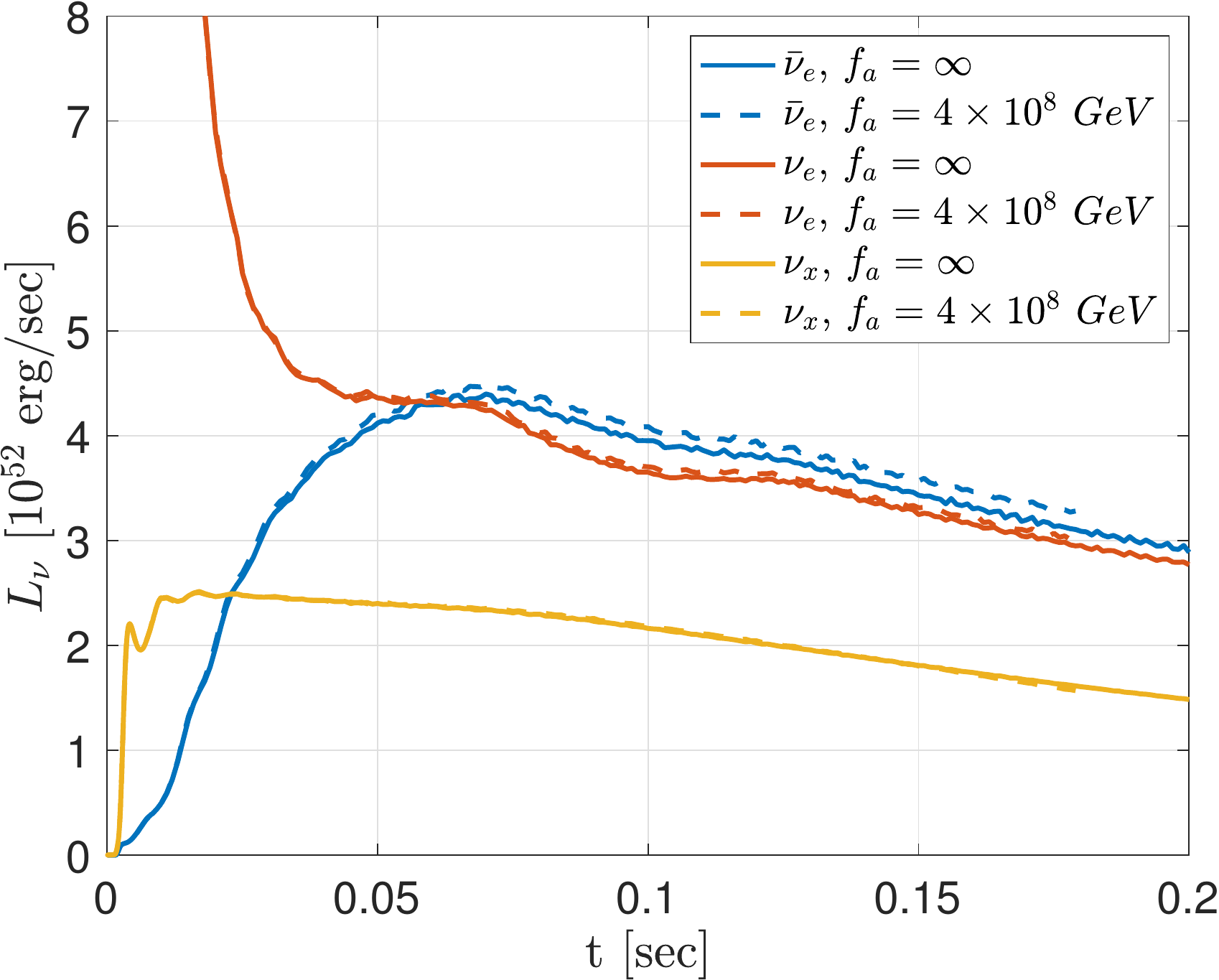}
\caption{Asymptotic neutrino luminosity calculated using GR1D with (dashed) and without (solid) axion emission.}
\label{fig:gr1d0}
\end{center}
\end{figure}

First, in Fig.~\ref{fig:gr1d0} we show the asymptotic neutrino luminosity vs post-bounce time, calculated with (dashed lines) and without (solid lines) axion emission. We can see that the asymptotic neutrino luminosity is unaffected by axion emission until the end of the simulation, at post-bounce time $t\sim0.2$~sec, corresponding to $\sim0.5$~sec after the onset of core collapse. 

\begin{figure}[htbp]
	\begin{center}
		\includegraphics[width=0.475\textwidth]{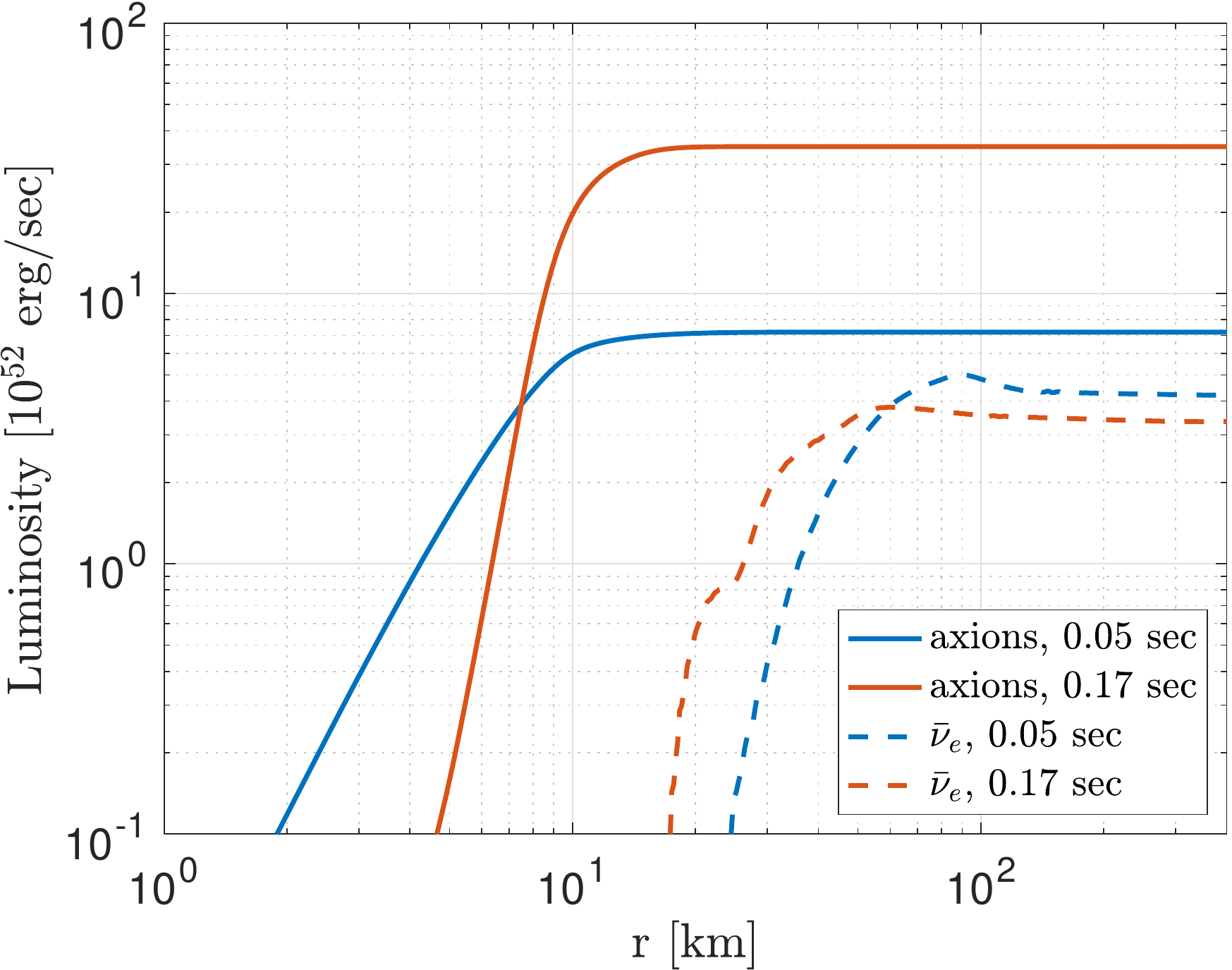}
		\caption{Axion and neutrino luminosity profiles as a function of radius from the centre of the star for two post-bounce time snapshots.}
		\label{fig:gr1dlum}
	\end{center}
\end{figure}

The asymptotic neutrino luminosity is unaffected by the axion emission not because the axion luminosity is small -- the axion luminosity is large in this simulation, as we show momentarily -- but because the neutrinos come from surface emission, and while the axions deplete the PNS core of energy, it takes a neutrino diffusion time (of order seconds) for this information to reach the surface. We demonstrate this point in Fig.~\ref{fig:gr1dlum}, where we show the radial profiles of the neutrino and axion luminosity build-up at two time snapshots, 0.05 and 0.17~sec post-bounce. The asymptotic axion luminosity grows from $\sim7\times10^{52}$~erg/s to $\sim3.5\times10^{53}$~erg/s from one snapshot to the second, in both cases reaching its asymptotic value inside of the PNS core at a radius of $\sim10$~km. In both snapshots the asymptotic axion luminosity is significantly larger than the neutrino luminosity that is seen to emerge from larger distances of order $r\sim50$~km. 

\begin{figure}[htbp]
	\begin{center}
		\includegraphics[width=0.475\textwidth]{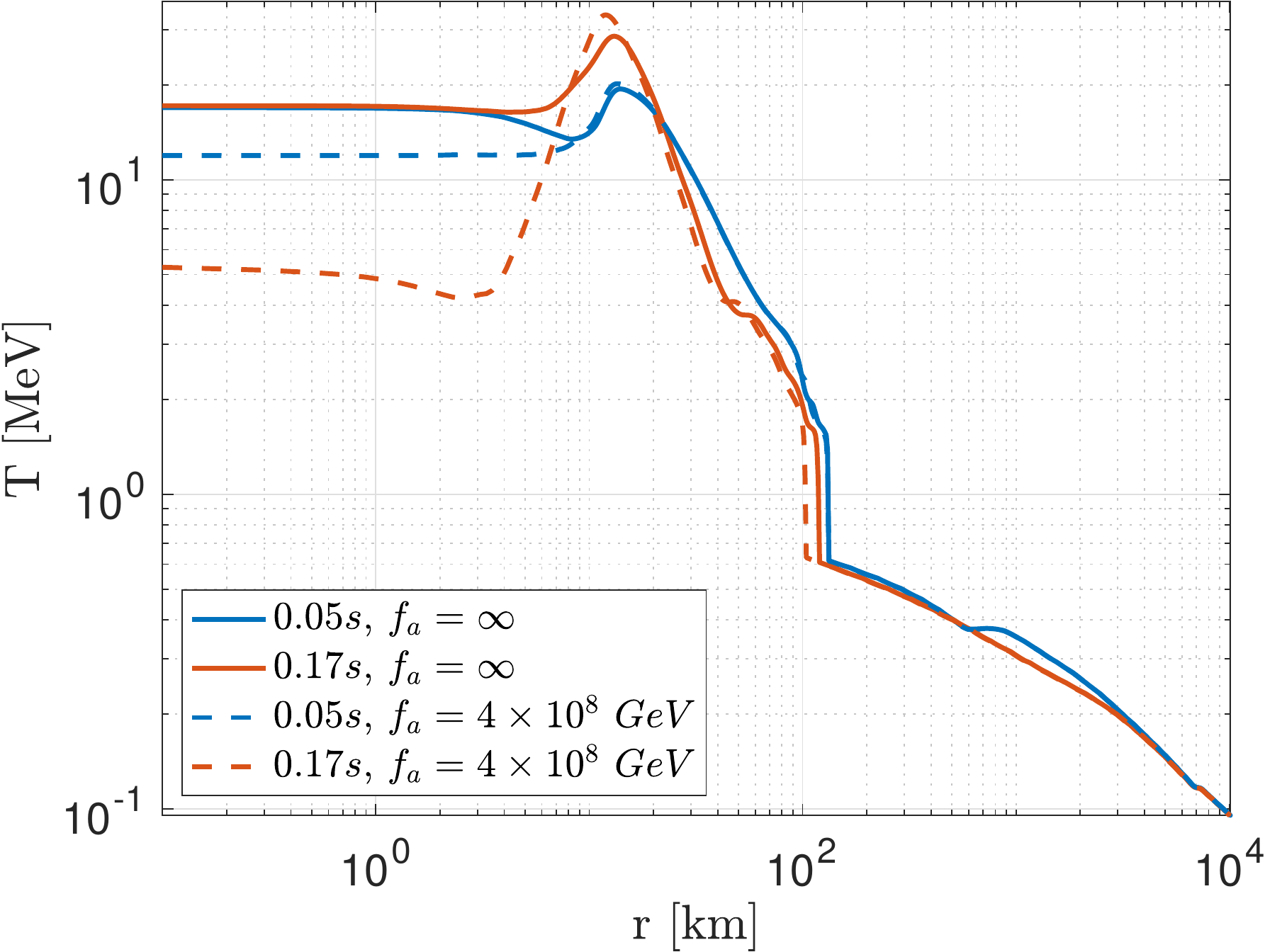}
		\includegraphics[width=0.475\textwidth]{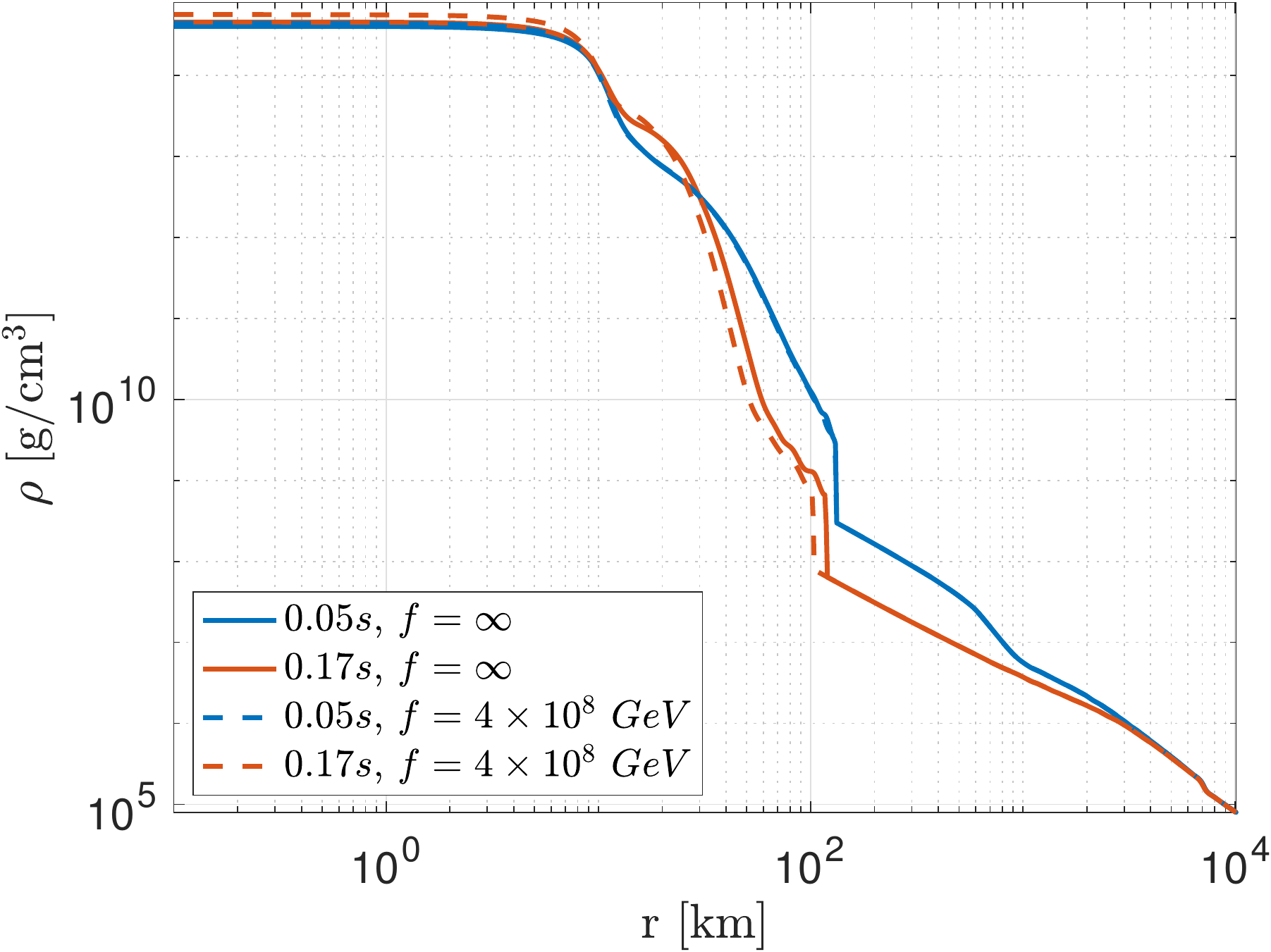}
		\includegraphics[width=0.475\textwidth]{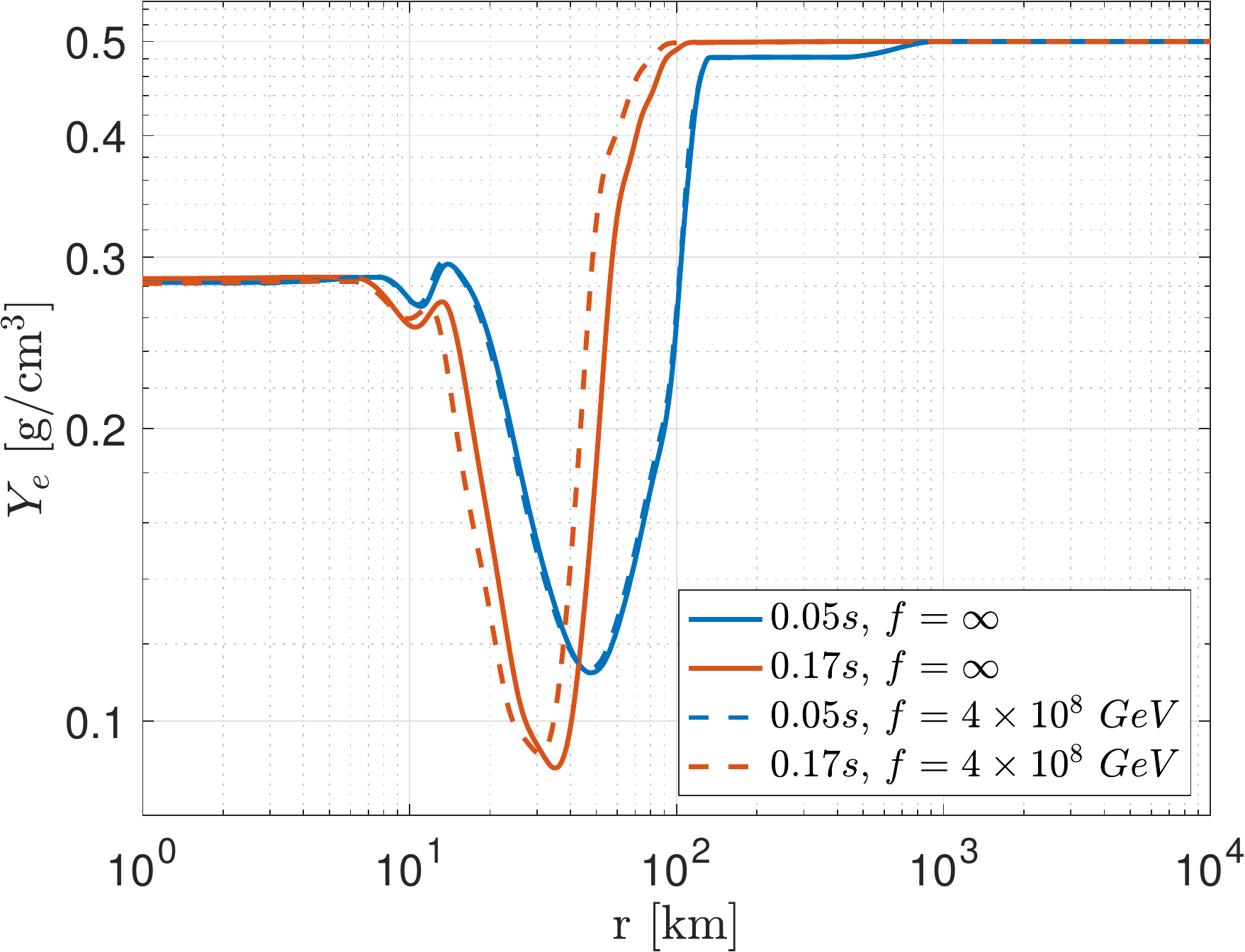}
		\caption{Temperature, density and electron fraction 
		profiles for different post-bounce time snapshots.}
		\label{fig:gr1d2}
	\end{center}
\end{figure}

In Fig.~\ref{fig:gr1d2} we demonstrate that the axion emission is indeed draining the PNS of its internal energy despite the fact that this effect is not manifest in neutrinos. In the top panel of Fig.~\ref{fig:gr1d2} we show the radial profile of the PNS temperature. In the axion emission simulation, the core temperature at $r\lesssim6$~km drops by a factor of $\sim3$ compared to the no-axion case by $t=0.17$~sec post-bounce, indicating large depletion of internal energy. Nevertheless, the PNS temperature profile at and above the neutrinosphere, $r\gtrsim20$~km, is essentially unaffected, explaining the stable neutrino signal. For completeness, in the bottom panels of Fig.~\ref{fig:gr1d2} we also show the PNS density and electron fraction in the different simulations.
\end{appendix}

\bibliography{ref}
%\begin{thebibliography}{99}
%\end{thebibliography}

\end{document}